\pdfoutput=1

\documentclass[aps,prl,twocolumn,showpacs]{revtex4}
\usepackage{graphicx}
\usepackage{amsmath,amssymb}
\usepackage{color}

\def\sign{\mathop{\rm sgn}}

\def\bk{{\bf k}}

\def\bq{{\bf q}}
\def\bQ{{\bf Q}}
\def\br{{\bf r}}

\input{epsf}


\begin{document}

\title{Conductivity close to antiferromagnetic criticality}

\author{S.V.~Syzranov and J.~Schmalian}
\affiliation{Institute for Theoretical Condensed Matter Physics,
  Karlsruhe Institute of Technology,
  76131 Karlsruhe, Germany}

\begin{abstract}
	We study the conductivity of a 3D disordered metal
	close to the antiferromagnetic instability within the framework of the spin-fermion model
	using the diagrammatic technique. We calculate the interaction correction
	$\delta\sigma(\omega,T)$ to the conductivity,
	assuming that the latter is dominated by the disorder scattering, and the interaction is weak.
	Although the fermionic scattering rate shows critical behaviour on the entire Fermi surface,
	the interaction correction is dominated by the processes near the hot spots, narrow regions of the
	Fermi-surface corresponding to the strongest spin-fermion scattering. Exactly at the critical
	point $\delta\sigma\propto[max(\omega, T)]^{3/2}$.
	At sufficiently large frequencies $\omega$
	the conductivity is independent of the temperature, and
	$\delta\sigma\propto(\tau^{-1}-i\omega)^{-2}$, $\tau$ being the elastic
	scattering time. In a certain intermediate frequency range
	$\delta\sigma(\omega)\propto i\omega(\tau^{-1}-i\omega)^{-2}$.
\end{abstract}


\maketitle

Transport close to quantum criticality fascinates and challenges
researchers in many fields of condensed matter, ranging from the physics of
high-temperature superconductors and heavy-fermion materials to the
conduction in graphene or granulated superconductors
near the superconductor-insulator transition\cite{Sachdev:book, vLohneysen:review}.
At low
temperatures, the interplay of disorder and interactions inevitably plays
an important role in transport. Efforts to investigate the
conductivity near magnetic instabilities often rely on the spin-fermion
model\cite{Abanov:review}: the conductivity is determined by low-energy
fermionic excitations, interacting with collective bosonic spin modes that
carry no charge yet become soft modes at the magnetic critical point.

If the interaction corrections to the conductivity are small, they can be
analysed in the framework of the spin-fermion model using perturbation
theory. For instance, for a 2D metal near a ferromagnetic (FM) instability
the corrections can be found microscopically%
\cite{Paul:FMconductivity} using the diagrammatic technique similar to the
electron-electron interaction corrections\cite%
{AltshulerAronov,Zala:intcorrection} to the conductivity of a disordered
metal. \ A separate analysis is needed in the case of an antiferromagnetic (AFM) instability,
characteristic of pnictide superconductors, certain heavy-fermion materials
and possibly the cuprate systems and organic charge transfer salts. Near the
AFM transition, the momenta of the lowest-energy spin fluctuations are large
and close to the reciprocal vectors of the spin superlattice in the AFM
phase. As a result, electrons strongly interact with spin fluctuations only
close to narrow regions of the Fermi surface, the so-called
\textquotedblleft hot spots\textquotedblright , and can be
scattered from one such region to another.

In a system without disorder the
quasiparticle lifetime and weight vanish at the hot spots at low energies.
The analysis of the Boltzmann transport equation concluded\cite{Hlubina:kinequation},
however, that the conductivity is dominated by weakly scattering
\textquotedblleft cold spots\textquotedblright. 
The role of impurities within the kinetic-equation approach 
was investigated by Ueda\cite{Ueda:magnetoresisance} and
Rosch\cite{Rosch:rubbish} who found a correction $\Delta
\rho \propto T^{3/2}$ to the residual resistivity $\rho
_{0}$ caused by impurities. The fractional power here was explained by the width of
the ``hot lines'' $\propto T^{1/2}$ \cite{Rosch:rubbish}. It is, however, unclear why such
a quasiclassical Boltzmann approach should be applicable, considering,
in particular, the
singular scattering rate and vanishing quasiparticle weight. Another interesting question,
which deserves a separate investigation, is
related to the notion of impurity-induced "local criticality"\cite{Si:lc}:
the single-particle self-energy may become momentum independent $\Sigma \left( 
\mathbf{k,}i\omega \right) \simeq \Sigma \left( i\omega \right)$
and show singular behaviour on the entire Fermi surface away from the hot spot,
that would manifest itself in the resistivity $\rho(T)$.
As estimated in Ref.~\cite{Woelfle:localcrit}, such
local scattering processes lead to the quasiparticle scattering time
$\tau_{sp}^{-1}\left( T\right)\propto T^{3/2}$. This momentum-independent
scattering rate implies a similar temperature dependence of $\Delta\rho$.
The emergence of an impurities-induced local single-particle scattering rate
is particularly interesting given the experimental indications in favour
of local, i.e. momentum independent, criticality of electrons near the
antiferromagnetic quantum critical points\cite{vLohneysen:review}.

In this Letter we study microscopically transport in the spin-fermion model
in the presence of impurities using the diagrammatic technique. We demonstrate that
at sufficiently strong disorder and weak spin-fermion coupling the interaction
corrections to the resistivity are dominated by processes near the hot spots.
The quasiparticle self-energies in the ``cold regions'' also show critical behaviour,
leading, however, to a smaller contribution to the resistivity correction.
We find the conductivity dependency on temperature and frequency.
In the zero-frequency low-temperature limit we recover the results previously known
from the kinetic-equation approach.

\begin{figure}[htbp]
	\centering
	\includegraphics[width=0.2\textwidth]{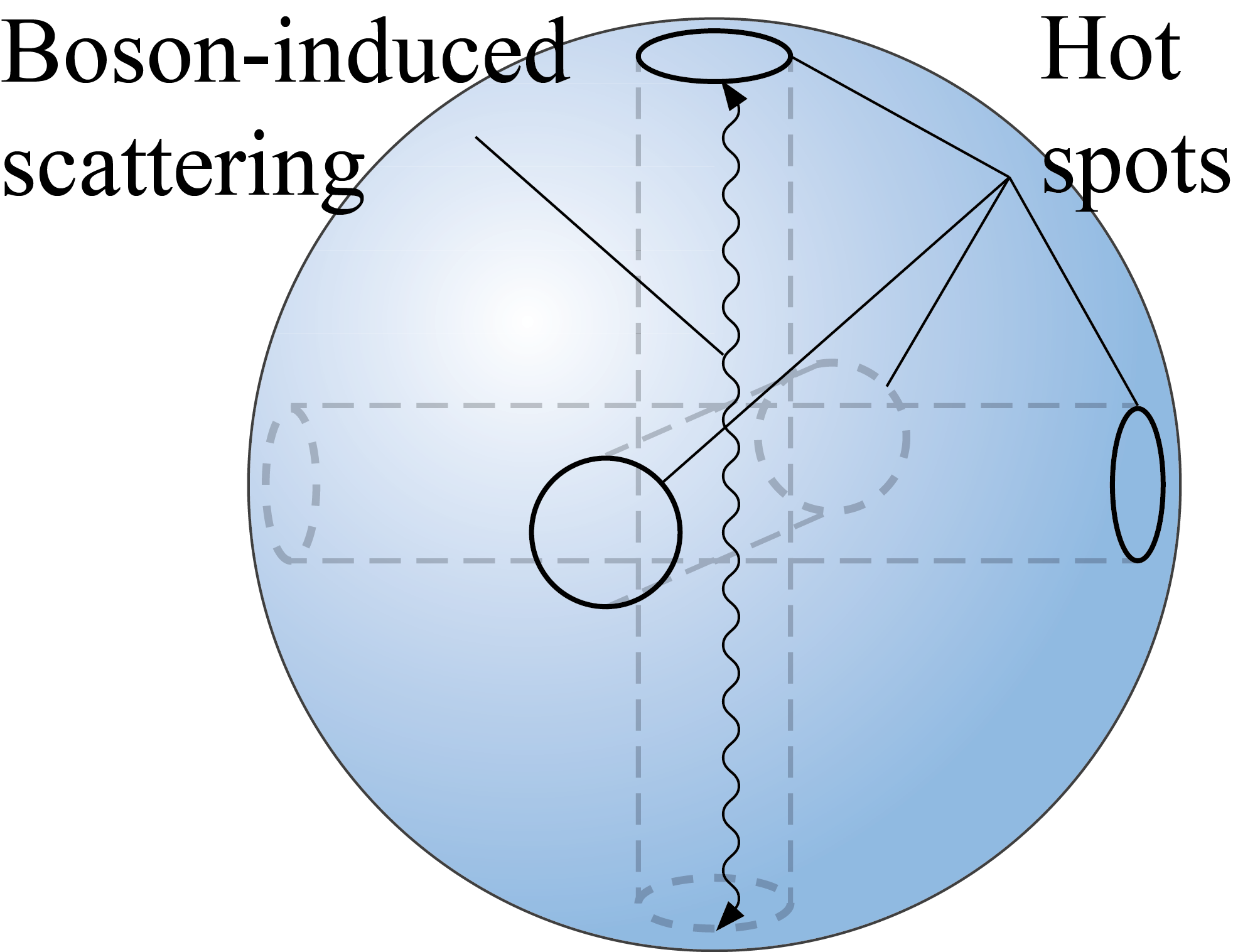}
	\caption{(Colour online) The Fermi surface.}
	\label{FermiS}
\end{figure}

{\it Model.} We consider for simplicity a spherical Fermi surface.
A fermion can absorb or emit only a boson with momentum
close to 
a certain large vector $\bQ$ ($|\bQ|\sim k_F$), determined by the
geometry of the AFM spin superlattice in the AFM phase.
In order to address the experimentally relevant case of an {\it isotropic} conduction
we assume that there are three such (incommensurate) vectors, $\bQ_1$, $\bQ_2$, and $\bQ_3$, of equal length
and directed respectively along different coordinate axes.
The fermions strongly interact with bosons only close to the 
``hot-spots'', points on the Fermi surface separated by the vectors $\bQ_n$,
Fig.~\ref{FermiS}. In our case the hot spots are three pairs of circles;
an electron near each circle can be inelastically scattered to the corresponding
circle in the opposite hemisphere.

The action of the spin-fermion model reads
\begin{eqnarray}
S =\sum_{\sigma}\int_{x}{c}_{\sigma }^{\dagger }(x)\left[ \partial _{\tau
}+\xi \left( -i\nabla \right) +U(\mathbf{r})\right] {c}_{\sigma }(x)
\nonumber \\
-\frac{1}{2}\sum_n\int_{x,x^{\prime }}\mathbf{S}_{n}(x){D_{n}^{-1}(x-x^{\prime })%
}\mathbf{S}_{n}(x^{\prime })  \nonumber \\
+{g}\sum_{\sigma \sigma ^{\prime }n}\int_{x}{c}_{\sigma }^{\dagger }\left(
x\right) \mathbf{\boldsymbol{\sigma }}_{\sigma \sigma ^{\prime }}{c}_{\sigma
^{\prime }}\left( x\right) \mathbf{S}_{n}(x),
\end{eqnarray}
where ${c}_{\sigma }^{\dagger }\left( x\right) $ and ${c}_{\sigma }\left(
x\right) $ are the Grassman fields for an electron (fermion) with spin $\sigma $, $x=\left( \mathbf{r,}\tau \right)$;
$\mathbf{S}_{n}\left(x\right)$ is the bosonic field of the collective spin excitations;
$n=1,2,3$ labels a
pair of hot spot circles separated by the vector $\mathbf{Q}_{n}$, $g$ is
the coupling constant between the bosons and the fermions, $\xi _{\mathbf{k}%
}\equiv\xi(\bk)$ is the electron spectrum, ${D_{n}(x-x^{\prime })}$ is the propagator of
the bosonic modes; the disorder is represented by the random potential $U(%
\mathbf{r})$, which acts on the fermions only. The fermion spectrum allows 
for the existence of the hot spots, i.e., points on the Fermi surface where $\xi_{\bk+\bQ}=\xi_\bk$.
We use the abbreviation $\int_{x}\cdots =\int d^{3}r\int_{0}^{\beta }d\tau \cdots $ and set $e=\hbar=1$.
Also, below we
suppress the bosonic index $n$, if the respective expression does not depend
on it, and use conventions $k=\left(\mathbf{k},i\omega\right)$ and
$\int_{k}\cdots =T\sum_\omega\int \frac{d\bk}{(2\pi)^3}$.

The energies of all excitations in the spin-fermion model are limited by a phenomenological scale
$\Lambda$, which separates the low energies involved in the transport phenomena from the high-energy modes
responsible for the formation of the antiferromagnetism.
The shape of the Fermi surface and the excitation spectra are assumed
renormalised upon having integrated out all the higher-energy modes
$\Lambda<\varepsilon\lesssim W$ \cite{Abanov:review}, where $W\gg\Lambda$
is the microscopic bandwidth of the tight-binding Hamiltonian
of the underlying lattice.
The relative smallness of the cutoff $\Lambda$ allows one to linearise
the fermionic spectrum $\xi_\bk$ with respect to $\mathbf{k-k}_{\mathrm{F}}$.

We consider a system sufficiently close to the critical point, so that the
collective spin excitations are the most important bosonic excitations in
the model and one can disregard the other types of interaction. On the other
hand, to address small interaction corrections to the conductivity of a
disordered metal, observed in experiments, we assume that the dimensionless
coupling constant, $\alpha \simeq {g^{2}}/{v_{F}}\ll 1$,
is the very smallest parameter in the theory,
which allows us
to treat the interactions perturbatively. 
In principle, the renormalised bosonic propagator $D_{n}(i\Omega ,\mathbf{q})$ depends on the coupling $g$.
However, we assume below that the boson dynamics is characterised by
the elastic scattering time $\tau$ of electrons and phenomenological energy scales independent
of $g$.

The assumption of small $\alpha$ implies, in particular, that the
conductivity is dominated by disorder scattering. For simplicity, the
impurity scattering is assumed to be isotropic, and the disorder potential
-- weak and Gaussian; $\langle U(\mathbf{r})U(\mathbf{r}^{\prime })\rangle
=(2\pi \rho_{F}\tau)^{-1}\delta(\br-\br^\prime)$, where $\rho_{F}=\frac{k_F^2}{2\pi^2v}$ is
the density of states on the Fermi surface. The parameter  
$(\varepsilon_F\tau)^{-1}\ll 1$
is assumed to be the second smallest in the problem, $\varepsilon_F$ being the Fermi energy.

{\it Perturbation theory.}  Under the assumptions that the bosonic modes are correlated on
a short scale yet the spin-fermion coupling is weak, one can conveniently
calculate the interaction corrections to the conductivity perturbatively. To
the $0^{th}$ order in the interaction, the conductivity is given
by the Drude contribution $\sigma _{0}\left(\omega \right)
=({2}/{3})v^{2}\rho_{F}\left(\tau^{-1}-i\omega \right)^{-1}$, together with the
weak-localisation and the electron-electron interaction corrections to it.
Next we calculate the leading correction $\sigma_{2}\propto g^2$ 
in the spin-fermion coupling $g$.

\begin{figure}[htbp]
	\centering
	\includegraphics[width=0.4\textwidth]{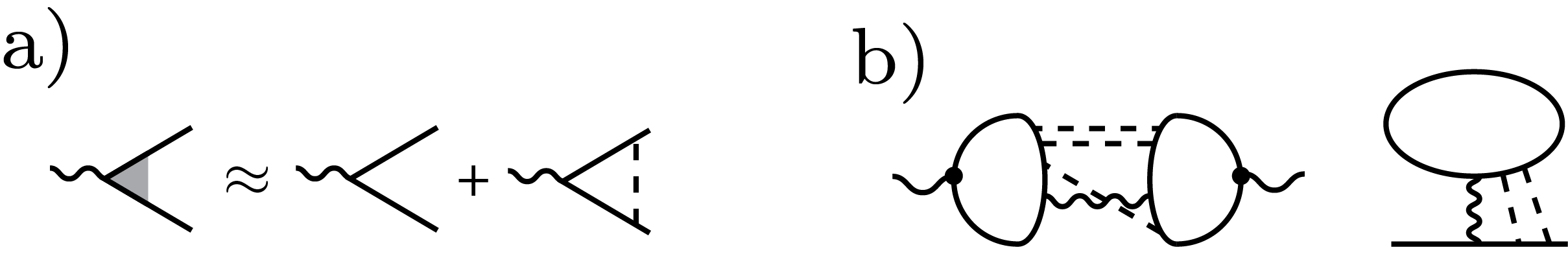}
	\caption{a)Renormalisation of the spin-fermion interaction vertex. 
	b)Hartree contributions to the conductivity and to the fermion scattering rate.}
	\label{VertexHartree}
\end{figure}

Let us consider first the renormalisation of the fermion-boson interaction
vertex by disorder. The first non-vanishing correction to the coupling $g$,
Fig.~\ref{VertexHartree}, reads 
\begin{equation}
	\delta g(\bq,i\omega,i\varepsilon)
	=\frac{g}{2\pi\rho_F\tau}\int\frac{d\bk}{(2\pi)^3} G(\mathbf{k},i\varepsilon )G(\mathbf{k}+%
\mathbf{q},i\varepsilon +i\omega),
	\label{grenorm}
\end{equation}
where $i\omega $ and $\mathbf{q}\simeq \mathbf{Q}$ are, respectively, the
frequency and the momentum of the spin excitation at the vertex; $%
G(i\varepsilon ,\mathbf{k})=\left[ i\varepsilon -\xi _{\mathbf{k}}+i/(2\tau)\sign\varepsilon\right]^{-1}$
is the disorder-averaged fermion propagator\cite{AGD}.
Because there are no excitations with energies larger than $\Lambda$,
the integration is confined to narrow regions around the hot-spot circles,
such that $|\xi_{\mathbf{k}},\xi_{\bk+\bq}|<\Lambda$.
This allows one to
integrate separately the two propagators near the two different hot-spot
circles, arriving at a very small renormalisation of the coupling
$\delta g/g\sim (\varepsilon _{F}\tau )^{-1}\sign\left(
\omega \right) \sign\left( \omega +\varepsilon \right)$,
which, as we show below, may be neglected when calculating the conductivity.
 Due to the large
momentum transfer by the AFM bosonic fluctuations, $|\bq|\sim|\mathbf{Q}|\sim
k_{F}$, the renormalisation is fully perturbative,
i.e. may be considered only to the leading order in the disorder potential,
and does not involve the summation of the whole diffusion ladder, unlike the
case of electron-electron or FM interactions, when the momentum scattering
at the vertex is small\cite{AltshulerAronov, Zala:intcorrection}.

Let us notice also that the Hartree contributions to
the conductivity and to the quasiparticle scattering rates, that contain one
boson propagator, (e.g., in Fig.~\ref{VertexHartree}b) vanish due to the
symmetry of the spin-fermion interaction vertex. Indeed, such diagrams
contain an independent summation of one interaction vertex with respect to
the spin polarizations $\propto \sum_{\sigma }\boldsymbol{\sigma }_{\sigma
\sigma ^{\prime }}=0$.

\begin{figure}[htbp]
	\centering
	\includegraphics[width=0.35\textwidth]{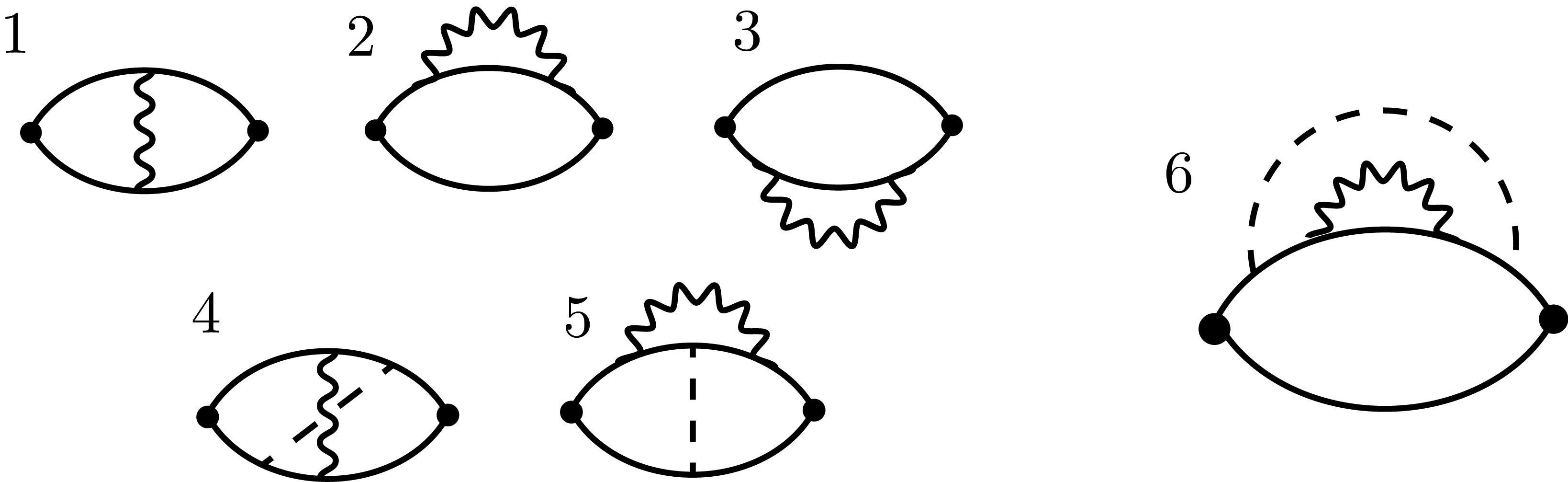}
	\caption{Interaction corrections to the conductivity.
	Diagrams 1-5 mimic processes close to the hot spots, while 6
	is the contribution of electrons on the whole Fermi surface.}
	\label{CondDiagrams}
\end{figure}

To the lowest order in interactions and in disorder strength the
conductivity correction is given by diagrams 1-3 in Fig.~\ref%
{CondDiagrams} (thin solid lines correspond to the disorder-averaged
electron propagators). Their contribution to the conductivity is given by the analytic
continuation from the Matsubara to real frequencies, $i\omega\rightarrow\omega+i0$,
of the quantity
\begin{equation}
\delta \sigma (i\omega )=\frac{\alpha v_{\parallel}k_F\int_{q}(\omega
-|\Omega |)\theta \left( \omega -|\Omega |\right) D(i\Omega,\bq)}
{\pi\omega {(\tau^{-1}+\omega )^{2}}},  \label{genericcond}
\end{equation}%
where $q=\left(\mathbf{q}, i\Omega\right)$ and $v_\parallel$ is the component
of velocity at the hot spot parallel to the respective vector $\bQ$.
The analytic continuation reads
\begin{equation}
\delta \sigma (\omega )=-\frac{2i\alpha v_\parallel k_F\int D^{R}(\varepsilon ,%
\mathbf{q})A(\omega ,\varepsilon) \frac{d\varepsilon d^{3}q}{\left(2\pi\right) ^{4}}}
{\pi\omega{(\tau^{-1}-i\omega )^{2}}},
\label{D}
\end{equation}
where $D^{R}(\varepsilon ,\mathbf{q})$ is the retarded boson propagator, and
$A\left( \omega ,\varepsilon \right) =(\omega
-\varepsilon )\left[ n(\varepsilon )-n(\varepsilon -\omega )\right] $ with the
Bose distribution function $n(\varepsilon )$.

Diagrams 4-6 in Fig.~\ref{CondDiagrams} contain one extra impurity line and
represent the next-order corrections to the conductivity in the disorder
strength, which are not accounted for by the renormalisation of the
interaction vertex. We show in what immediately follows that, because of
the large bosonic momenta, adding more
impurity lines to diagrams 1-3 results in small corrections, that are
perturbative in the disorder strength.

The straightforward evaluation of the diagram 4 shows that the corresponding
contribution $\delta \sigma _{4}(i\omega )\sim \delta \sigma (i\omega )%
\mathcal{O}\left[ (\varepsilon_{F}\tau)^{-1}\right]$ is suppressed by the small
parameter $(\varepsilon_{F}\tau)^{-1}$ compared to diagrams 1-3, which can be
understood as follows. The extra impurity line in diagram 4 adds to the
respective integral one more momentum integration and two propagators, whose
momenta are shifted by a large constant vector of the order of $k_{F}$ with
respect to each other. This extra integration results in the relative
smallness $(\varepsilon_{F}\tau)^{-1}$ of the diagram. A similar argument
proves the smallness of the diagrams with crossing impurity lines in the
usual disorder-averaging diagrammatic technique\cite{AGD}. Let us emphasise
that this smallness is specific of the case to the large momentum scattering
by the bosonic modes close to AFM criticality. If the electron dynamics
is strongly affected by collective excitations with small momenta, e.g., FM
fluctuations\cite{Paul:FMconductivity} or electron-electron interactions\cite%
{AltshulerAronov}, then taking into account the whole diffusion ladder is
necessary in place of the impurity lines for diagrams 4 and 5, as well
as in the renormalisation of the spin-fermion interaction vertex.

Similarly, the contribution of diagram 5, 
\begin{equation}
\delta \sigma _{5}(i\omega )\sim \alpha\tau^{-1}(\tau^{-1}+\omega )^{-2}
\int_q D(i\Omega ,\mathbf{q})
\label{diagr5}
\end{equation}
is also suppressed by the small parameter $(\varepsilon_{F}\tau)^{-1}$, compared to
diagrams 1-3. However, Eq.~(\ref{diagr5}) involves the summation of the boson propagator
over all Matsubara frequencies, unlike Eq.~(\ref{genericcond}), where the frequency summation
is restricted to $|\Omega|<|\omega|$.
This difference can make diagram 5 important at sufficiently low frequencies
$\omega$. However, the temperature
dependencies of both diagrams are determined by a few terms with
$\Omega\sim T$ in the sums in Eqs.~(\ref{diagr5}) and (\ref{genericcond}),
and we can neglect the temperature dependency contribution of the diagram 5
so long as $\varepsilon_F\tau\gg1$. 
$\delta\sigma_5(\omega=0,T=0)$ amounts to a contribution to the residual
conductivity due to the spin fluctuations
and may be disregarded in what follows.
Similarly, one can neglect the diagrams for the conductivity that correspond to
the renormalisation of the interaction vertex since $\varepsilon_F\tau\gg1$.

\begin{figure}[htbp]
	\centering
	\includegraphics[width=0.25\textwidth]{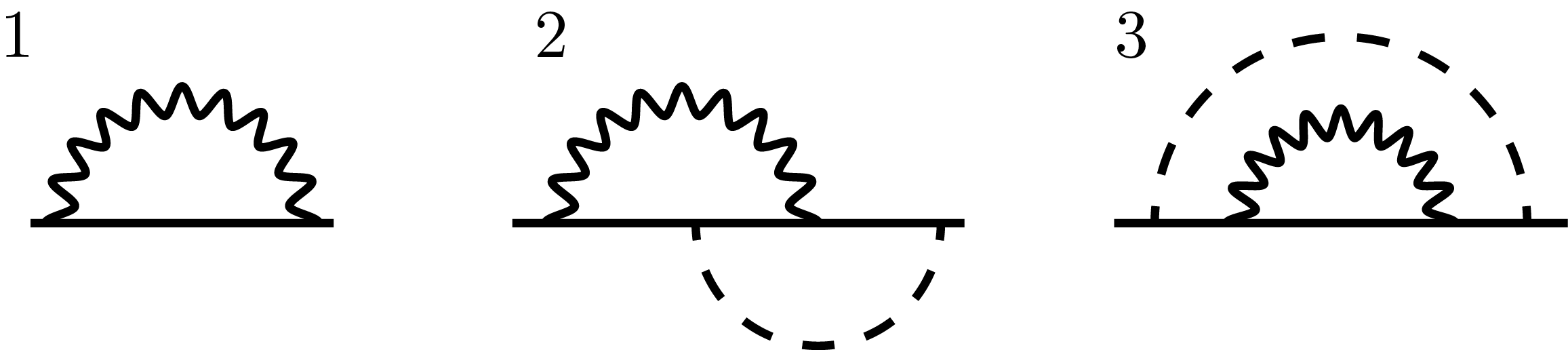}
	\caption{Interaction corrections to the scattering rate. Diagrams 1 and 2
	describe scattering close to the hot spots, while 3 is an isotropic contribution
	to the scattering on the whole Fermi surface.}
	\label{Srate}
\end{figure}

Let us proceed to diagram 6 in Fig.~\ref{CondDiagrams}.
This interaction correction comes from electrons near the whole Fermi surface, unlike
the previously considered diagrams 1-5, which correspond to processes near the hot
spots. Indeed, anywhere on the Fermi surface an electron can be scattered into a
hot spot, where its dynamics is impeded by spin fluctuations, cf. diagram 3 in
Fig.~\ref{Srate}. The value of the diagram 6 in Fig.~\ref{CondDiagrams}
\begin{eqnarray}
	\delta\sigma_6(i\omega)\sim(\Lambda\tau)^{-1}\delta\sigma(i\omega)
	\label{rainbowsmall}
\end{eqnarray}
is small compared to the value $\delta\sigma(i\omega)$ of diagrams 1-3 due to the
small parameter $(\Lambda\tau)^{-1}\ll1$, which can be understood as follows.
On one hand, the extra smallness, $\sim(\varepsilon_F\tau)^{-1}$, comes from the
extra impurity line. On the other hand, the diagram contains an extra largeness
in it, $\sim\varepsilon_F/\Lambda$,-- the ratio of the area of the Fermi surface to the
characteristic size of the hot spots. The combination of the two factors leads
to Eq.~(\ref{rainbowsmall}).

Thus, we have shown that to the leading order in disorder and interactions the conductivity
is given by Eq.~(\ref{genericcond}), corresponding to diagrams 1-3 in Fig.~\ref{CondDiagrams}.
To the first order in $\alpha$ the conductivity reads
\begin{eqnarray}
	\sigma \left( \omega \right) =\frac{2}{3}v^2\rho_F
	\left[{\tau^{-1}+\delta\tau^{-1}
	-i\omega \left( 1+\lambda\right) }\right]^{-1},
\end{eqnarray}
where we introduced
\begin{eqnarray}
	\delta\tau^{-1}=-\frac{6\pi\alpha}{k_F \omega}\frac{v_\parallel}{v}
	\int{Im}D^R(\varepsilon,\bq)A(\omega,\varepsilon)\frac{d\varepsilon d\bq}{(2\pi)^4},
	\label{deltatau}
	\\
	\lambda=-\frac{6\pi\alpha}{k_F\omega^2}\frac{v_\parallel}{v}
	\int{Re}D^R(\varepsilon,\bq)A(\omega,\varepsilon)\frac{d\varepsilon d\bq}{(2\pi)^4}.
	\label{lambda}
\end{eqnarray}

{\it The limit of low frequencies and temperatures.} To make further progress at sufficiently
small $\omega$ and $T$ we consider the most general form of the overdamped boson propagator that
follows from the effective Ginzburg-Landau-type description:
\begin{equation}
	D^R_n(\varepsilon,\bq)=-\sum_\pm\left[{\xi^{-2}+(\bq\pm\bQ)^2-i\varepsilon k_F^2\gamma^{-1}}\right]^{-1}
	\label{bosonpropagator}
\end{equation}
at $|\varepsilon|\ll \varepsilon_F^2\gamma^{-1}$. For $|\varepsilon|\gg \varepsilon_F^2\gamma^{-1}$,
$D_n^R\propto\varepsilon^{-2}$. The scale $\gamma$ characterises elastic
and inelastic scattering of the bosons. If the latter is attributable to electron-hole pairs,
$\gamma=4\pi vv_\parallel k_F g^{-2}$. However, being a parameter of the boson propagator,
$\gamma$ should be regarded as an independent phenomenological scale in the spin-fermion model
and may come, e.g., from disorder scattering.

Substituting Eq.~(\ref{bosonpropagator}) into Eq.~(\ref{deltatau}) yields
$\delta\tau^{-1}=\frac{3\alpha\zeta(3/2)}{4\sqrt{2\pi}}\frac{v_\parallel}{v}
\frac{T^{3/2}}{\gamma^{1/2}}$ for $(k_F\xi)^2\gg\gamma T^{-1}$ and $\omega=0$.
While at $(k_F\xi)^2\ll\gamma T^{-1}$ and $\omega=0$ it holds that
$\delta\tau^{-1}=\frac{\pi\alpha}{2}\frac{v_\parallel}{v}\frac{T^2k_F\xi}{\gamma}$.
Similarly, for $T=0$ and finite frequencies $\delta\tau^{-1}=\frac{\alpha\sqrt{2}}{5\pi}
\frac{v_\parallel}{v}\frac{\omega^{3/2}}{\gamma^{1/2}}$. The crossover
between these regimes is described by the expression
\begin{eqnarray}
	\delta\tau^{-1}=\alpha\frac{v_\parallel}{v}\frac{T^\frac{3}{2}}{\gamma^\frac{1}{2}}
	\Phi\left(\frac{\gamma}{(k_F\xi)^2T},\frac{\omega}{T}\right)
\end{eqnarray}
with the scaling function
\begin{equation}
	\Phi(y,z)=\frac{3}{2\pi^2z}\int
	\frac{x(z-x)t^2\left(\frac{1}{e^x-1}-\frac{1}{e^{x-z}-1}\right)}
	{(y+t^2)^2+x^2}dx dt.
\end{equation}
In particular, in the dc limit the correction to the resistivity due to the
spin fluctuations at the AFM critical point reads
\begin{equation}
	\delta\rho=\alpha C\frac{v_\parallel T^\frac{3}{2}}{v^2k_F^2\tau^2\gamma^\frac{1}{2}},
\end{equation}
where $C=\frac{9\pi^\frac{3}{2}\zeta(3/2)}{4\sqrt{2}}=23.14..$, which up to a numerical coefficient
matches the result obtained within the kinetic-equation approach\cite{Ueda:magnetoresisance, Rosch:rubbish}.

{\it High frequencies.} 
At larger $\omega$ the frequency dependency of the conductivity can be found
regardless of the particular form of the boson propagator. If the frequency is very large,
the cutoff $\Lambda$ of the excitation energies in the spin-fermion model
should be chosen larger than $\omega$ to ensure the existence
of quasiparticles that can absorb a quantum $\omega$.

Let us assume first that the bosonic dynamics is determined by certain energy
scales independent of the cutoff $\Lambda$. Then one can neglect the bosonic
frequencies $\Omega$ in comparison with the large $\omega$ in Eq.~(\ref{genericcond}).
Making the analytic continuation and introducing the average value of the spin fluctuations
$\langle S^2\rangle=-\int_q D(i\Omega,\bq)$, we arrive at
\begin{eqnarray}
	\delta\sigma(\omega)=-\frac{\alpha v_\parallel k_F}{\pi (\tau^{-1}-i\omega)^2}
	\langle S^2\rangle.
	\label{condhigh}
\end{eqnarray}

The result (\ref{condhigh}) has a simple physical interpretation.
At very high frequencies the spin fluctuations may be considered
frozen and equivalent to static disorder. The appropriate modification
of the elastic scattering time, averaged with respect to angles, estimates
$\delta\langle 1/\tau\rangle\sim \alpha\langle S^2\rangle/k_F$
and causes a correction $\delta\sigma\sim(\delta\tau^{-1})\partial\sigma_0/\partial\tau^{-1}$
to the Drude conductivity $\sigma_0$. As is necessary,
this correction matches Eq.~(\ref{condhigh}).

{\it Intermediate frequencies.} Let us proceed to the intermediate frequencies;
$\omega$ exceeds the temperature $T$, but is smaller than the smallest
characteristic energy of the boson dynamics, e.g., the bosonic scattering rate $\varepsilon_F^2\gamma^{-1}$.

Then at $|\Omega|<\omega$ the integral $\int D(i\Omega,\bq)\frac{d\bq}{(2\pi)^3}$ is nearly independent
of $\Omega$, because the value of $\Omega$ affects the boson propagator only on a sufficiently small momentum
interval,
while the rest of the integral
is accumulated on a greater interval.

Thus, in Eq.~(\ref{genericcond}) we may set $\int D(i\Omega,\bq)(d\bq)\approx\int D(0,\bq)(d\bq)$.
Then the frequency summation in Eq.~(\ref{genericcond}) and the analytic continuation to real
frequencies yield
\begin{equation}
	\delta\sigma(\omega)\propto-i\omega(\tau^{-1}-i\omega)^{-2},
	\label{condmod}
\end{equation}
independently of the form of the bosonic propagator.

{\it Discussion.} Spin-fermion interactions modify the quasiparticle self-energy part
on the whole Fermi surface. Away from the hot spots the modification
\begin{equation}
	\Sigma_{lc}(i\varepsilon)\sim-i\alpha(k_F\Lambda\tau)^{-1}
	\int_q \theta(|\varepsilon|-|\Omega|)D(i\Omega,\bq)\sign\varepsilon.
	\label{Sigmalc}
\end{equation} is given
by diagram 3 in Fig.~\ref{Srate}.
It is momentum-independent and
``locally critical'', corresponding to the scattering rate
$\tau_{lc}^{-1}\propto const+T^{3/2}$. Let us notice that Eqs.~(\ref{deltatau}), (\ref{lambda}),
and (\ref{Sigmalc}) hold for an arbitrary boson propagator and can be used to analyse
transport in, e.g., a system with a two-dimensional spin dynamics\cite{vLohneysen:review}.
In arbitrary dimensions $d$, it holds that $\tau_{lc}^{-1}\propto const+T^{d/2}$.

We demonstrated, however, that the interaction correction to the conductivity
near the AFM instability is dominated by the processes near the hot spots, corresponding to
diagrams 1-3 in Fig.~\ref{CondDiagrams}. We have found the dependency of the conductivity on frequency $\omega$
and temperature $T$. In the limit $\omega=0$ we recover the temperature dependencies 
of the interaction correction to the conductivity previously known
from the kinetic equation analysis; $\delta\sigma\propto T^{3/2}$ and $\delta\sigma\propto T^2$
at the critical point and away from it respectively. At $T=0$ we find $\delta\sigma\propto\omega^{3/2}$.
At sufficiently high frequencies the correction is independent of temperature and a particular form of
the spin propagator. At very high and moderate frequencies the dependency is given by Eqs.~(\ref{condhigh})
and (\ref{condmod}), respectively.

{\it Acknowledgements.}
We appreciate useful discussions with B.N.~Narozhny
and P.~W{\"o}lfle.


\begin{thebibliography}{12}
\expandafter\ifx\csname natexlab\endcsname\relax\def\natexlab#1{#1}\fi
\expandafter\ifx\csname bibnamefont\endcsname\relax
  \def\bibnamefont#1{#1}\fi
\expandafter\ifx\csname bibfnamefont\endcsname\relax
  \def\bibfnamefont#1{#1}\fi
\expandafter\ifx\csname citenamefont\endcsname\relax
  \def\citenamefont#1{#1}\fi
\expandafter\ifx\csname url\endcsname\relax
  \def\url#1{\texttt{#1}}\fi
\expandafter\ifx\csname urlprefix\endcsname\relax\def\urlprefix{URL }\fi
\providecommand{\bibinfo}[2]{#2}
\providecommand{\eprint}[2][]{\url{#2}}

\bibitem[{\citenamefont{Sachdev}(2011)}]{Sachdev:book}
\bibinfo{author}{\bibfnamefont{S.}~\bibnamefont{Sachdev}},
  \emph{\bibinfo{title}{Quantum Phase Transitions}}
  (\bibinfo{publisher}{Cambridge University Press}, \bibinfo{year}{2011}).

\bibitem[{\citenamefont{v.~L{\"o}hneysen
  et~al.}(2007)\citenamefont{v.~L{\"o}hneysen, Rosch, Vojta, and
  W{\"o}lfle}}]{vLohneysen:review}
\bibinfo{author}{\bibfnamefont{H.}~\bibnamefont{v.~L{\"o}hneysen}},
  \bibinfo{author}{\bibfnamefont{A.}~\bibnamefont{Rosch}},
  \bibinfo{author}{\bibfnamefont{M.}~\bibnamefont{Vojta}}, \bibnamefont{and}
  \bibinfo{author}{\bibfnamefont{P.}~\bibnamefont{W{\"o}lfle}},
  \bibinfo{journal}{Rev. Mod. Phys.} \textbf{\bibinfo{volume}{79}},
  \bibinfo{pages}{1015} (\bibinfo{year}{2007}).

\bibitem[{\citenamefont{Abanov et~al.}(2002)\citenamefont{Abanov, Chubukov, and
  Schmalian}}]{Abanov:review}
\bibinfo{author}{\bibfnamefont{A.}~\bibnamefont{Abanov}},
  \bibinfo{author}{\bibfnamefont{A.~V.} \bibnamefont{Chubukov}},
  \bibnamefont{and}
  \bibinfo{author}{\bibfnamefont{J.}~\bibnamefont{Schmalian}},
  \bibinfo{journal}{Adv. Phys.} \textbf{\bibinfo{volume}{52}},
  \bibinfo{pages}{119} (\bibinfo{year}{2002}).

\bibitem[{\citenamefont{Paul et~al.}(2005)\citenamefont{Paul, Pepin, Narozhny,
  and Maslov}}]{Paul:FMconductivity}
\bibinfo{author}{\bibfnamefont{I.}~\bibnamefont{Paul}},
  \bibinfo{author}{\bibfnamefont{C.}~\bibnamefont{Pepin}},
  \bibinfo{author}{\bibfnamefont{B.~N.} \bibnamefont{Narozhny}},
  \bibnamefont{and} \bibinfo{author}{\bibfnamefont{D.~L.}
  \bibnamefont{Maslov}}, \bibinfo{journal}{Phys. Rev. Lett.}
  \textbf{\bibinfo{volume}{95}}, \bibinfo{pages}{017206}
  (\bibinfo{year}{2005}).

\bibitem[{\citenamefont{Altshuler and Aronov}(1985)}]{AltshulerAronov}
\bibinfo{author}{\bibfnamefont{B.~L.} \bibnamefont{Altshuler}}
  \bibnamefont{and} \bibinfo{author}{\bibfnamefont{A.~G.}
  \bibnamefont{Aronov}}, in \emph{\bibinfo{booktitle}{Electron-electron
  interactions in disordered systems}}, edited by
  \bibinfo{editor}{\bibfnamefont{A.~L.} \bibnamefont{Efros}} \bibnamefont{and}
  \bibinfo{editor}{\bibfnamefont{M.}~\bibnamefont{Pollak}}
  (\bibinfo{publisher}{North-Holland}, \bibinfo{address}{Amsterdam},
  \bibinfo{year}{1985}).

\bibitem[{\citenamefont{Zala et~al.}(2001)\citenamefont{Zala, Narozhny, and
  Aleiner}}]{Zala:intcorrection}
\bibinfo{author}{\bibfnamefont{G.}~\bibnamefont{Zala}},
  \bibinfo{author}{\bibfnamefont{B.}~\bibnamefont{Narozhny}}, \bibnamefont{and}
  \bibinfo{author}{\bibfnamefont{I.}~\bibnamefont{Aleiner}},
  \bibinfo{journal}{Phys. Rev. B} \textbf{\bibinfo{volume}{64}},
  \bibinfo{pages}{214204} (\bibinfo{year}{2001}).

\bibitem[{\citenamefont{Hlubina and Rice}(1995)}]{Hlubina:kinequation}
\bibinfo{author}{\bibfnamefont{R.}~\bibnamefont{Hlubina}} \bibnamefont{and}
  \bibinfo{author}{\bibfnamefont{T.~M.} \bibnamefont{Rice}},
  \bibinfo{journal}{Phys. Rev. B} \textbf{\bibinfo{volume}{51}},
  \bibinfo{pages}{9253} (\bibinfo{year}{1995}).

\bibitem[{\citenamefont{Ueda}(1977)}]{Ueda:magnetoresisance}
\bibinfo{author}{\bibfnamefont{K.}~\bibnamefont{Ueda}}, \bibinfo{journal}{J.
  Phys. Soc. Japan} \textbf{\bibinfo{volume}{43}}, \bibinfo{pages}{1497}
  (\bibinfo{year}{1977}).

\bibitem[{\citenamefont{Rosch}(1999)}]{Rosch:rubbish}
\bibinfo{author}{\bibfnamefont{A.}~\bibnamefont{Rosch}},
  \bibinfo{journal}{Phys. Rev. Lett.} \textbf{\bibinfo{volume}{82}},
  \bibinfo{pages}{4280} (\bibinfo{year}{1999}).

\bibitem[{\citenamefont{Si et~al.}(2001)\citenamefont{Si, Rabello, Ingersent,
  and Smith}}]{Si:lc}
\bibinfo{author}{\bibfnamefont{Q.}~\bibnamefont{Si}},
  \bibinfo{author}{\bibfnamefont{S.}~\bibnamefont{Rabello}},
  \bibinfo{author}{\bibfnamefont{K.}~\bibnamefont{Ingersent}},
  \bibnamefont{and} \bibinfo{author}{\bibfnamefont{J.~L.} \bibnamefont{Smith}},
  \bibinfo{journal}{Nature} \textbf{\bibinfo{volume}{413}},
  \bibinfo{pages}{804} (\bibinfo{year}{2001}).

\bibitem[{\citenamefont{W{\"o}lfle and Abrahams}(2011)}]{Woelfle:localcrit}
\bibinfo{author}{\bibfnamefont{P.}~\bibnamefont{W{\"o}lfle}} \bibnamefont{and}
  \bibinfo{author}{\bibfnamefont{E.}~\bibnamefont{Abrahams}},
  \bibinfo{journal}{Phys. Rev. B} \textbf{\bibinfo{volume}{84}},
  \bibinfo{pages}{041101(R)} (\bibinfo{year}{2011}).

\bibitem[{\citenamefont{Abrikosov et~al.}(1975)\citenamefont{Abrikosov, Gorkov,
  and Dzyaloshinski}}]{AGD}
\bibinfo{author}{\bibfnamefont{A.~A.} \bibnamefont{Abrikosov}},
  \bibinfo{author}{\bibfnamefont{L.~P.} \bibnamefont{Gorkov}},
  \bibnamefont{and} \bibinfo{author}{\bibfnamefont{I.~E.}
  \bibnamefont{Dzyaloshinski}}, \emph{\bibinfo{title}{Methods of Quantum Field
  Theory in Statistical Physics}} (\bibinfo{publisher}{Dover, New York},
  \bibinfo{year}{1975}).

\end{thebibliography}
\end{document}